# Dust-acoustic instability driven by drifting ions and electrons in the dust plasma with Lorentzian kappa distribution

Zhipeng Liu, Jiulin Du

*School of Science, Department of Physics, Tianjin University, Tianjin 300072, China*

**Abstract**

The instability of the dust-acoustic waves driven by drifting electrons and ions in a dusty plasma is investigated by the kinetic theory. All the plasma components (electrons, ions and dust particles) are assumed to be the Lorentzian $\kappa$ (kappa)-distributions. The spectral indexes $\kappa$ of the $\kappa$-distributions for the three plasma components are different from each other. The obtained instability growth rate depends on the physical quantities of the plasma not only, but on the spectral indexes. The numerical results for the $\kappa$-effect on the instability growth rate show that, if the normalized wave number is small, the index of electrons has a stabilized effect on the dust acoustic waves and the index of ions has an instability effect on the waves, but if the normalized wave number is large, they both nearly have no any effect on the waves. In reverse, the index of dust grains has nearly no any effect on the instability growth rate if the normalized wave number is small, but it has a stabilized effect on the dust waves if the normalized wave number is large.

**I. INTRODUCTION**

Non-Maxwellian distributions can be observed commonly both in space plasmas and in laboratory plasmas. For instance, The plasmas in the planetary magnetospheres are found to be deviation from Maxwellian distribution due to the presence of high energy particles.[1] The spacecraft measurements of plasma velocity distributions, both in the solar wind and in the planetary magnetospheres and magnetosheaths, have revealed that non-Maxwellian distributions are quite common. In many situations the distribution have a "suprathermal" power-law tail at high energies, which has been well modeled by the so-called Lorentzian $\kappa$ (kappa)-distribution.[2]

There are several possible distributions that have been constructed in order to fit the empirical data. Among these distributions, the family of kappa distribution has attracted much attention, for its interesting applications were found in the space and laboratory plasmas.[3,4] The $\kappa$-distribution has been extensively used to analyze a large number of phenomena in plasmas. For instance, Maksimovic et al[5] presented a kinetic model of solar wind based on the $\kappa$-velocity distribution functions for the electrons and protons escaping out of the solar corona. Thorne and Summers[6] applied this distribution to the well known Landau damping phenomenon to study the electron plasma waves and ion acoustic waves in space plasmas. Recently, Baluku and Hellberg[7] investigated the dust acoustic solitons in dusty plasma with $\kappa$-distributed ions and/or electrons, and found that only negative potential solitons existed when the dust is negative. Zaheer et al[8] and Rolffs[9] el al used $\kappa$-distribution functions to study the anisotropic Weibel instability. Lee[10] used the $\kappa$-distribution to investigate the Landau damping of dust-acoustic waves. Meanwhile, the applications of the $\kappa$-distribution in the space plasmas have long needed to find the foundation of statistical theory. Recently, the $q$-distribution in nonextensive statistical mechanics has been applied to investigate the non-Maxwellian distribution in the plasmas such as the ion-acoustic waves,[11,12] the solar interior,[13] the solar wind,[14,15] and the interplanetary space plasma.[16] Therefore, the nonextensive statistical mechanics may provide one statistical foundation for the use of the kappa distribution. These results are opening a window of investigating electromagnetic modes in various space plasma scenarios where the non-Maxwellian distributions are relevant. There has been a growing interest in the non-Maxwellian distributions of the plasma components in plasma physics.

In this paper, using the kinetic theory, we investigate the instability of dust-acoustic waves driven by weak ion and electron drifts in the collisionless magnetic-field-free dusty plasma. We assume the plasma to be described by the Lorentzian $\kappa$-distribution for the electrons, ions and dust grains. Since ions, electrons and dust grains interact with each other via the different long-range interactions, including Coulombian forces and the gravitations,[13] which can be described by the spectral indexes of distribution functions for the plasma components,[17] we will investigate the instability of dust-acoustic waves given the different spectral $\kappa$-indexes for the distribution function. Thereby, the nonextensive effects on the three different components can be found respectively.

## II. INSTABILITY GROWTH RATE OF DUST ACOUSTIC WAVES

Conventionally, the three-dimensional $\kappa$-distribution function is given by[18]

$$f_0(\mathbf{v}) = \frac{n_0}{\pi^{3/2}\theta^3\kappa^{3/2}} \frac{\Gamma[\kappa+1]}{\Gamma[\kappa-1/2]} \left[1 + \frac{1}{\kappa}\frac{\mathbf{v}^2}{\theta^2}\right]^{-(\kappa+1)}, \tag{1}$$

where $\kappa$ is the spectral index, $n_0$ is particle number density, $k_B$ is Boltzmann's constant, $T$ and $m$ are the temperature and mass of the particles, $\theta$ is the effective thermal speed, related to the index $\kappa$ by

$$\theta = \left(\frac{2\kappa-3}{2\kappa}\frac{2k_BT}{m}\right)^{1/2}, \tag{2}$$

The Maxwellian distribution can be recovered form Eq.(1) in the limit $\kappa \to \infty$. Generally, the $\kappa$-distribution (1) should be subject to suprathermal tails for decreasing $\kappa$ values, but the thermal spread in Eq. (2) decreases for decreasing $\kappa$ values. Leubner[16,19] used a modified $\kappa$ distribution by replacing the thermal speed $\theta$ in Eq.(1) by the Maxwellian one $v_t = \sqrt{2k_BT/m}$,

$$f_0(\mathbf{v}) = \frac{n_0}{\pi^{3/2}v_t^3\kappa^{3/2}} \frac{\Gamma[\kappa+1]}{\Gamma[\kappa-1/2]} \left[1 + \frac{1}{\kappa}\frac{\mathbf{v}^2}{v_t^2}\right]^{-(\kappa+1)}. \tag{3}$$

Furthermore, when the power $-(\kappa+1)$ in Eq.(3) is replaced by $-\kappa$ and then if using the transformation $\kappa = 1/(q-1)$, Eq.(3) becomes the q-distribution in nonextensive statistics.

In this work, we use Eq.(1) as the basic undisturbed non-Maxwellian distribution function in the dust plasma to investigate the instability of the dust-acoustic waves. Our approach should also be suitable for the case using Eq.(3) or the q-distribution function in nonextensive statistics as the basic undisturbed non-Maxwellian distribution function.

We consider the collisionless, unbounded, magnetic-field-free dusty plasma with electrons, singly charged ions and negatively or positively charged dust grains. The dust grains are treated as an additional charged species of uniform mass $m_d$ and charge $Z_d$. The dust plasma is considered as in the statistical distribution described by Eq.(1). When the plasma departs slightly from the equilibrium state, the $\kappa$-distribution function can be approximately written as

$$f_\alpha = f_{\alpha 0}(v) + f_{\alpha 1}(\vec{v},\vec{r},t), \quad f_{\alpha 1} \ll f_{\alpha 0}, \tag{4}$$

where the subscript $\alpha$ refers to the components, with $\alpha = i, e$ and $d$ for the ions, the electrons and the dust, respectively. $f_{\alpha 0}$ is the $\kappa$-distribution of the components $\alpha$ and $f_{\alpha 1}$ is the perturbation about $f_{\alpha 0}$. The dynamical behavior of the dust plasma is governed by the Vlasov equations and the Poisson's equations, respectively,

$$\frac{\partial f_\alpha}{\partial t} + \mathbf{v} \cdot \nabla f_\alpha - \frac{Q_\alpha}{m} \nabla \phi \cdot \nabla_{\mathbf{v}} f_\alpha = 0, \tag{5}$$

$$\nabla^2 \phi = -\frac{1}{\varepsilon_0} \sum_\alpha Q_\alpha \int f_{\alpha 1} d\mathbf{v}, \tag{6}$$

where $\phi$ is the electrostatic potential, $Q_\alpha$ is the charge of the components $\alpha$. The charge neutrality condition is

$$\varepsilon_d Z_d n_{d0} + Z_i n_{i0} - n_{e0} = 0 \tag{7}$$

where $Z_d$ and $Z_i$ are the charge of the dust particle and the ion respectively; $\varepsilon_d = +1, (-1)$ for positively (negatively) charged dust grain. Throughout this paper, hydrogen ions are assumed and $Z_i = +1$. Without lose of generality, we assume the plane wave to propagate in the $z$-direction. Taking the perturbations in the form of $f_{\alpha 1} \sim \exp i(kz - \omega t)$, we make Fourier transformation for Eq. (5) and (6). Then the combination of them leads to the equation,

$$\varepsilon_\kappa(\omega, k) \phi = 0 \tag{8}$$

where $\varepsilon_\kappa(\omega, k)$ is the dielectric permittivity, given by

$$\varepsilon_\kappa(\omega, k) = 1 - \sum_\alpha \frac{\omega_{p\alpha}^2}{k^2} \frac{1}{n_{\alpha 0}} \int \frac{\partial f_{\alpha 0} / \partial v_z}{(v_z - \omega / k)} d^3v, \tag{9}$$

where $\omega_{p\alpha} = \sqrt{n_{\alpha 0} Q_\alpha^2 / \varepsilon_0 m_\alpha}$ is the plasma frequency. Since the electrostatic potential has non-trivial solution, then we have $\varepsilon_\kappa(\omega, k) = 0$, being the dispersion relation,

$$1 + \sum_\alpha \frac{\omega_{p\alpha}^2}{k^2} \frac{1}{n_{\alpha 0}} \int \frac{k \partial f_{\alpha 0} / \partial v_z}{\omega - k v_z} d^3v = 0. \tag{10}$$

Let the distribution functions for the electrons, ions and dust grains be the drifting Lorentzian $\kappa$-distributions,

$$f_{\kappa,\alpha}(\mathbf{v}) = \frac{n_{\alpha 0}}{\pi^{3/2} \theta_\alpha^3 \kappa_\alpha^{3/2}} \frac{\Gamma[\kappa_\alpha + 1]}{\Gamma[\kappa_\alpha - 1/2]} \left[ 1 + \frac{1}{\kappa_\alpha} \frac{v_x^2 + v_y^2 + (v_z - u_{\alpha 0})^2}{\theta_\alpha^2} \right]^{-(\kappa_\alpha + 1)}, \tag{11}$$

the electrons and ions are drifting together with $u_{e0} = u_{i0} = u_0$, and the dust grains are rest at $u_{d0} = 0$, then $u_0$ is the relative speed of the ion/electron to the dust grain. Inserting Eq.(11) into Eq.(10), we obtain

$$1 + \sum_{\alpha=i,e} \frac{1}{k^2 \lambda_{D\alpha}^2} \left( \frac{\kappa_\alpha}{\kappa_\alpha - 3/2} \right) \left[ \frac{\kappa_\alpha - 1/2}{\kappa_\alpha} + (\xi_\alpha - u_0') Z_\kappa(\xi_\alpha - u_0') \right]$$
$$+ \frac{1}{k^2 \lambda_{Dd}^2} \left( \frac{\kappa_d}{\kappa_d - 3/2} \right) \left[ \frac{\kappa_d - 1/2}{\kappa_d} + \xi_d Z_\kappa(\xi_d) \right] = 0, \tag{12}$$

where $\lambda_{D\alpha} = (k_B T_\alpha / m_\alpha)^{1/2} \omega_{p\alpha}^{-1}$ is the Debye length of the components $\alpha$, $\xi_\alpha = \omega / k \theta_\alpha$ is the phase velocity normalized to the thermal velocity $\theta_\alpha$, $u_0' = u_0 / \theta_\alpha$, and $Z_\kappa(\xi)$ is the modified plasma dispersion function,[21]

$$Z_\kappa(\xi) = \frac{\Gamma[\kappa + 1]}{\pi^{1/2} \kappa^{3/2} \Gamma[\kappa - 1/2]} \int_{-\infty}^{\infty} \frac{dx}{(x - \xi)(1 + x^2/\kappa)^{\kappa+1}}. \tag{13}$$

It is easy to find that in the limit $\kappa \to \infty$, the well-known Fried-Conte dispersion function is recovered

$$Z(\xi) = \frac{1}{\sqrt{\pi}} \int_{-\infty}^{\infty} \frac{e^{-x^2}}{x - \xi} dx. \tag{14}$$

For the dust acoustic waves, the phase velocity is much larger than the thermal velocity of the dust but much less than those of the ions and the electrons, namely, $v_{Td} \ll v_\phi \ll v_{Ti} \ll v_{Te}$. From the dimensionless parameter $\xi_\alpha = v_\phi / v_{T\alpha}$, one gets $\xi_e \ll \xi_i \ll 1$ and $\xi_d \gg 1$. Making the small argument expansion for the ions and electrons and the large argument expansion for dust grains,[20] one finds Eq.(12) to become

$$1+\sum_{\alpha=e,i}\frac{1}{k^2\lambda_{D\alpha}^2}\left(\frac{\kappa_\alpha}{\kappa_\alpha-3/2}\right)\left\{\frac{\kappa_\alpha-1/2}{\kappa_\alpha}+\frac{\omega-u_0}{k\theta_\alpha}\frac{\kappa_\alpha!\sqrt{\pi}i}{\kappa_\alpha^{3/2}\Gamma(\kappa_\alpha-1/2)}\right.$$

$$\left.-\frac{(2\kappa_\alpha-1)(2\kappa_\alpha+1)}{2\kappa_\alpha^2}\left(\frac{\omega-u_0}{k\theta_\alpha}\right)^2\right\}+\frac{1}{k^2\lambda_{Dd}^2}\left(\frac{\kappa_d}{\kappa_d-3/2}\right)$$

$$\times\left\{\frac{\kappa_d-1/2}{\kappa_d}+\frac{\omega}{k\theta_d}\frac{\kappa_d!\kappa_d^{\kappa_d-1/2}\sqrt{\pi}i}{\Gamma(\kappa_d-1/2)}\left(\frac{\omega}{k\theta_d}\right)^{-2(\kappa_d+1)}+\left(\frac{2\kappa_d-1}{2\kappa_d}\right)\left[1+\left(\frac{\kappa_d}{2\kappa_d-1}\right)\left(\frac{\omega}{k\theta_d}\right)^{-2}\right]\right\}.$$

(15)

The real part and imaginary part of Eq.(15) are, respectively,

$$\varepsilon_\kappa^{re}(\omega,k)=1+\sum_{\alpha=e,i}\frac{1}{k^2\lambda_{D\alpha}^2}\left(\frac{2\kappa_\alpha-1}{2\kappa_\alpha-3}\right)-\frac{1}{k^2\lambda_{Dd}^2}\left(\frac{\omega}{k\theta_d}\right)^{-2}\left(\frac{2\kappa_d}{2\kappa_d-3}\right) \quad (16)$$

and

$$\varepsilon_\kappa^{im}(\omega,k)=\sum_{\alpha=e,i}\frac{1}{k^2\lambda_{D\alpha}^2}\left(\frac{\kappa_\alpha}{\kappa_\alpha-3/2}\right)\left(\frac{\omega-u_0}{k\theta_\alpha}\frac{\kappa_\alpha!\sqrt{\pi}i}{\kappa_\alpha^{3/2}\Gamma(\kappa_\alpha-1/2)}\right)$$

$$+\frac{1}{k^2\lambda_{Dd}^2}\left(\frac{\kappa_d}{\kappa_d-3/2}\right)\left(\frac{\omega}{k\theta_d}\frac{\kappa_d!\kappa_d^{\kappa_d-1/2}\sqrt{\pi}i}{\Gamma(\kappa_d-1/2)}\left(\frac{\omega}{k\theta_d}\right)^{-2(\kappa_d+1)}\right),$$

(17)

where $\varepsilon_\kappa^{re}(\omega,k)$ and $\varepsilon_\kappa^{im}(\omega,k)$ are respectively the real part and imaginary part of the dielectric permittivity $\varepsilon_\kappa(\omega,k)$. Let the wave frequency be complex $\omega=\omega_r+i\gamma_\kappa$, where $\omega_r$ and $\gamma_\kappa$ are the real and imaginary parts of $\omega$, respectively, and let $\varepsilon_\kappa^{re}=0$, we find

$$\frac{\omega_r}{\omega_{pd}}=k\lambda_{De}\left(k^2\lambda_{De}^2+\frac{2\kappa_e-1}{2\kappa_e-3}+\delta\frac{T_e}{T_i}\frac{2\kappa_i-1}{2\kappa_i-3}\right)^{-1/2}, \quad (18)$$

and

$$\gamma_\kappa=-\frac{\varepsilon_\kappa^{im}(\omega,k)}{\partial\varepsilon_\kappa^{re}(\omega,k)/\partial\omega}\bigg|_{\omega=\omega_r}. \quad (19)$$

where $\delta=n_{i0}/n_{e0}$ is the ion-electron density ratio. Eq.(18) is the generalized dispersion relation of the dust acoustic waves defined in the $\kappa$-distribution. As $\kappa_\alpha\to\infty$, the standard dispersion relation in Maxwellian distribution can be recovered form Eq.(18). Substituting Eq.(16) and (17) into Eq.(19), one has the generalized instability growth rate of the dust acoustic waves,

$$\frac{\gamma_\kappa}{\omega_{pd}} = A_\kappa \left[ B_\kappa \left(1 - D_\kappa \frac{u_0}{v_{Te}}\right) + C_\kappa \right], \quad (20)$$

where

$$A_\kappa = -\frac{k\lambda_{De}\sqrt{\varepsilon_d Z_d (1-\delta)}}{2} \sqrt{\frac{\kappa_e}{2\kappa_e - 3}} \left( k^2 \lambda_{De}^2 + \frac{2\kappa_e - 1}{2\kappa_e - 3} + \delta \frac{T_e}{T_i} \frac{2\kappa_i - 1}{2\kappa_i - 3} \right)^{-2},$$

(21)

$$B_\kappa = \frac{\kappa_e}{\kappa_e - 3/2} \frac{\sqrt{\pi}\kappa_e!}{\kappa_e^{3/2} \Gamma(\kappa_e - 1/2)} \left(\frac{m_e}{m_d}\right)^{1/2}$$
$$+ \sqrt{\frac{\kappa_e - 3/2}{\kappa_e}} \frac{\sqrt{\pi}\kappa_i!}{(\kappa_i - 3/2)^{3/2} \Gamma(\kappa_i - 1/2)} \delta \left(\frac{m_i}{m_d}\right)^{1/2} \left(\frac{T_e}{T_i}\right)^{3/2}$$

(22)

$$C_\kappa = \sqrt{\frac{\kappa_e - 3/2}{\kappa_e}} \frac{\sqrt{\pi}\kappa_d!}{(\kappa_d - 3/2)^{3/2} \Gamma(\kappa_d - 1/2)} \varepsilon_d Z_d (1-\delta) \left(\frac{T_e}{T_d}\right)^{3/2}$$
$$\times \left[ 1 + \frac{T_e/T_d}{2\kappa_d - 3} \varepsilon_d Z_d (1-\delta) \left( k^2 \lambda_{De}^2 + \frac{2\kappa_e - 1}{2\kappa_e - 3} + \delta \frac{T_e}{T_i} \frac{2\kappa_i - 1}{2\kappa_i - 3} \right)^{-1} \right]^{-(\kappa_d + 1)},$$

(23)

and

$$D_\kappa = \frac{\sqrt{2}}{\sqrt{\varepsilon_d Z_d (1-\delta)}} \sqrt{\frac{m_d}{m_e}} \left( k^2 \lambda_{De}^2 + \frac{2\kappa_e - 1}{2\kappa_e - 3} + \delta \frac{T_e}{T_i} \frac{2\kappa_i - 1}{2\kappa_i - 3} \right)^{1/2}.$$

(24)

The instability occurs if $\gamma_\kappa > 0$, which is equivalent to $u_0 > u_0^{th}$ (the threshold drift velocity), so from Eq.(20) we have

$$u_0 > v_{Te} \frac{1}{D_\kappa} \left(1 + \frac{C_\kappa}{B_\kappa}\right) \quad (25)$$

As seen from Eq.(20), the generalized instability growth rate of the dust acoustic waves depends not only on the physical quantities, such as the mass, the temperature, the density ratios of the components, and the dust charge state, but also on the spectral indexes $\kappa_\alpha$. In the limit $\kappa_\alpha \to \infty$, $\gamma_\kappa$ reduces to the growth rate in Maxwellian distribution,[21]

$$\gamma = -\sqrt{\frac{\pi}{8}}\sqrt{\varepsilon_d Z_d(1-\delta)}\frac{k\lambda_{De}}{\left(1+k^2\lambda_{De}^2+\delta T_e/T_i\right)^2}\left\{\left[\left(\frac{m_e}{m_d}\right)^{1/2}+\delta\left(\frac{m_i}{m_d}\right)^{1/2}\left(\frac{T_e}{T_i}\right)^{3/2}\right]\right.$$

$$\times\left[1-\left(1+k^2\lambda_{De}^2+\delta\frac{T_e}{T_i}\right)\frac{\sqrt{2}}{\sqrt{\varepsilon_d Z_d(1-\delta)}}\frac{u_0}{v_{Te}}\right]+\varepsilon_d Z_d(1-\delta)\left(\frac{T_e}{T_d}\right)^{3/2}$$

$$\left.\times\exp\left[-\frac{T_e}{2T_d}\frac{\varepsilon_d Z_d(1-\delta)}{\left(1+k^2\lambda_{De}^2+\delta T_e/T_i\right)}\right]\right\}.$$

(26)

If let $u_0 = 0$ in Eq.(20), we find the Landau damping of the dust acoustic waves in the $\kappa$-distribution,

$$\gamma_\kappa^{da} = A_\kappa(B_\kappa + C_\kappa),\tag{27}$$

If there is a simple case of $\kappa_e = \kappa_i = \kappa$ and $\kappa_d \to \infty$, namely, let the ions and the electrons be in the Lorentzian distribution with the same index $\kappa$, but the dust grains be in the Maxwellian distribution, then Eq.(27) reduces to the result by Lee,[10]

$$\gamma_\kappa^{da} = -\frac{\sqrt{\pi}\kappa\Gamma(\kappa)\omega_{pd}^2 k^2\lambda_{Di}^2\left(1+\sqrt{\frac{m_e T_i^3}{m_i T_e^3}}\frac{1}{\delta}\right)\frac{2\kappa-3}{2\kappa-1}}{2\Gamma(\kappa+1/2)(\kappa-3/2)^{1/2}\left(\frac{2\kappa-3}{2\kappa-1}k^2\lambda_{Di}^2+1+\frac{T_i}{T_e}\frac{1}{\delta}\right)^2 kv_{Ti}}.\tag{28}$$

### III. NUMERICAL RESULTS AND DISCUSSIONS

In this section, by the numerical technique, we investigate the effects of the indexes, $\kappa_\alpha\,(\alpha=e,i,d)$, on the instability growth rate in the dust plasma with the $\kappa$-distributed ions, electrons and dust grains. As shown from Eq.(20), the instability growth rates depend on the quantities, such as the mass, the temperature, the density ratios between the particles of the components, the dust charge state, and the drifting-phase velocity ratio. Although these quantities in space plasmas are quite uncertain, the nominal values are given in some space environments.[22-24] For example, in the Saturn's E-ring or G-ring, the nominal values[23,24] are: $Z_d \sim 7\times 10^4$, $\delta = n_i/n_e \simeq 1.004$. We let $T_e = 2T_i = 10T_d$ and $v_{re}/v_{Te} = 1/100$. With this values, we can regard the normalized instability growth rate, $\gamma_\kappa/\omega_{pd}$, as a function of the normalized wavenumber, $k\lambda_{De}$, and plot in the figures 1, 2, 3, respectively, the effects of $\kappa_e$, $\kappa_i$ and $\kappa_d$ on the wave growth rate.

Figure 1 shows the effect of $\kappa_e$ on the growth rate, the calculation is based on Eq.(20), where we have let $\kappa_i = \kappa_d = \infty$. We find that instability growth rate grows with the increase of $\kappa_e$ and has the maximum at $\kappa_e = \infty$ (Maxwellian distribution). So we conclude that the $\kappa$-distribution of the electrons in the plasma has a stabilized effect on the dust acoustic waves if the normalized wave number $k\lambda_{De}$ is small, but it nearly has no any effect if the normalized wave number $k\lambda_{De}$ is large. The dust acoustic waves will be more unstable for a bigger $\kappa_e$ than for a smaller $\kappa_e$.

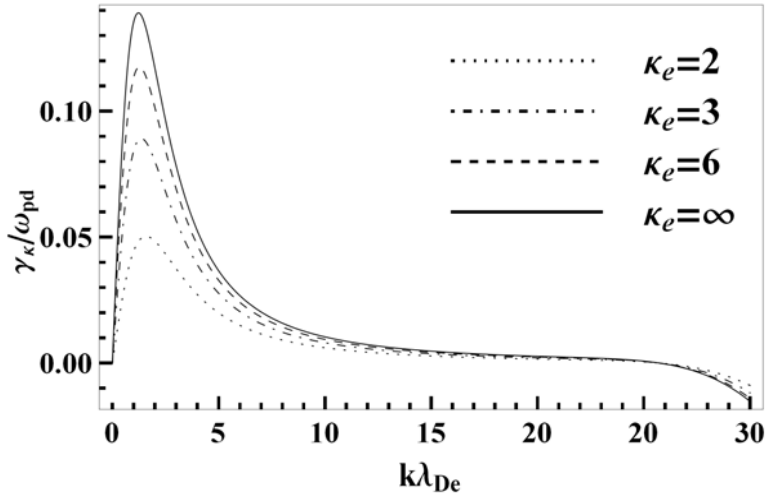

Fig. 1 Normalized instability growth rate of dust acoustic waves as a function of the normalized wave number $k\lambda_{De}$ for different values of $\kappa_e$.

Figure 2 shows the effect of $\kappa_i$ on the growth rate, the calculation is based on Eq.(20), where we have let $\kappa_e = \kappa_d = \infty$. We find that instability growth rate drops with the increase of $\kappa_i$ and has the minimum at $\kappa_i = \infty$ (Maxwellian distribution), thus the dust acoustic waves will become more stable with the increase of $\kappa_i$. So we conclude that the $\kappa$-distribution of the ions in the plasma has an instability effect on the dust acoustic waves if the normalized wave number $k\lambda_{De}$ is small, but if the normalized wave number $k\lambda_{De}$ is large, it nearly has no any effect on the rate. The dust acoustic waves will be more unstable for a smaller $\kappa_i$ than with a bigger $\kappa_i$.

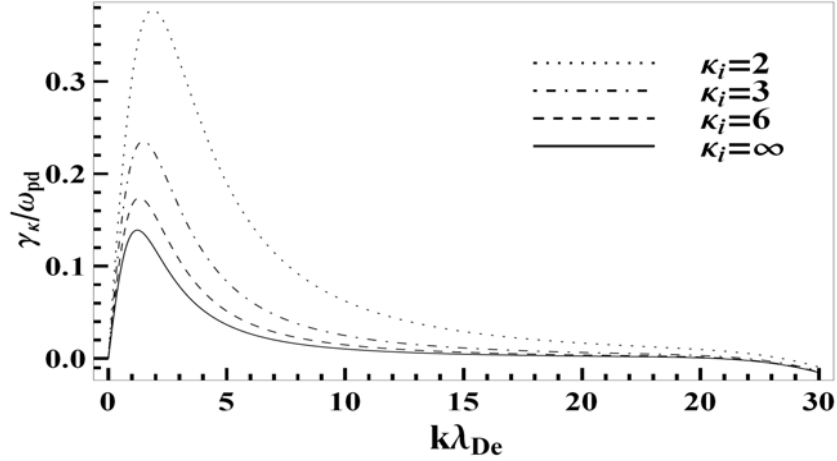

Fig. 2. Normalized instability growth rate of dust acoustic waves as a function of the normalized wave number $k\lambda_{De}$ for the different values of $\kappa_i$.

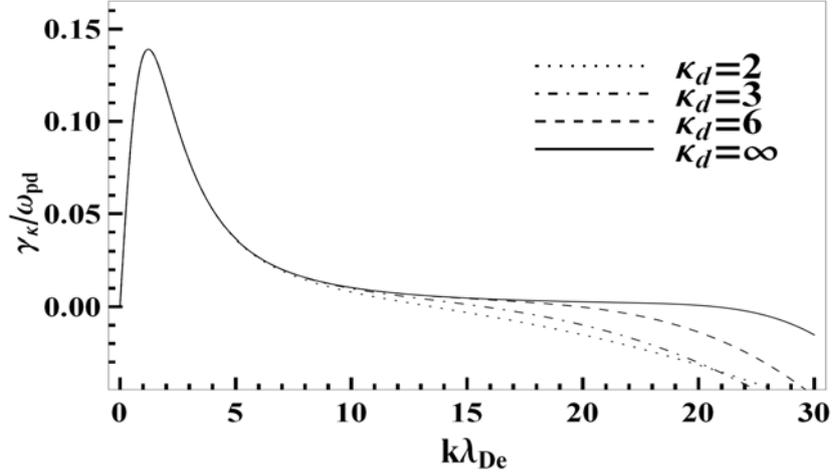

FIG. 3. Normalized instability growth rate of dust acoustic waves as a function of the normalized wave number $k\lambda_{De}$ for the different values of $\kappa_d$

Figure 3 shows the effect of $\kappa_d$ on the growth rate, the calculation is based on Eq.(20), where we have let $\kappa_i = \kappa_e = \infty$. We find that if the normalized wave number $k\lambda_{De}$ is small, the $\kappa$-distribution of the dust grains in the plasma has nearly no any effect on the instability growth rate of the dust acoustic waves, but if the normalized wave number $k\lambda_{De}$ is large, the instability growth rate increases with the increase of $\kappa_d$ and has the maximum at $\kappa_d = \infty$ (Maxwellian distribution). Thus we conclude that the $\kappa$-distribution of the dust grains in the plasma has a stabilized effect on the dust acoustic waves, the dust acoustic waves will become more stable for a deviation of the dust grains from Maxwellian distribution.

If we let $\kappa_e = \kappa_i = \kappa$ and $\kappa_d = \infty$ in Eq.(20), we can investigate the joint effects on the instability growth rate. Figure 4 shows that, in this case, the $\kappa$-distribution of the ions and the electrons has an instability effect on the dust acoustic waves in the plasma. The deviation of the ions and the electrons from Maxwellian distribution will lead to the increase of the instability growth rate of the dust acoustic waves.

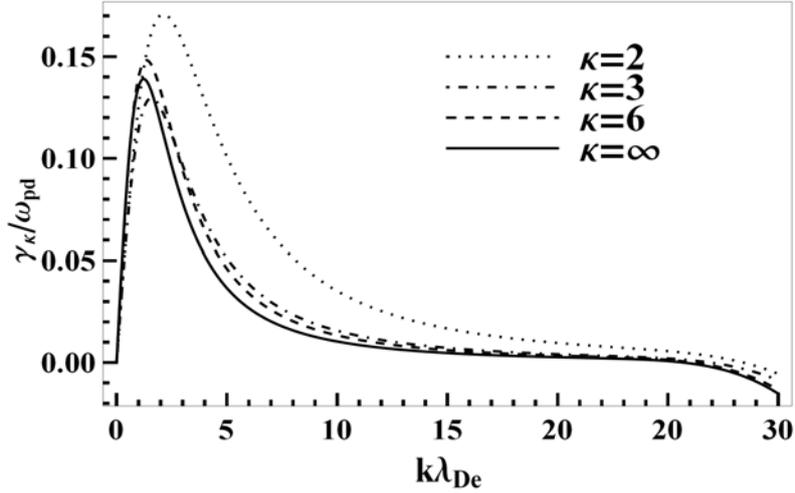

FIG. 4. Normalized instability growth rate of dust acoustic waves as a function of the normalized wave number $k\lambda_{De}$ in the case of $\kappa_e = \kappa_i = \kappa$ and $\kappa_d = \infty$.

## IV. SUMMARY and CONCLUSION

We have investigated the instability growth rate of dust-acoustic waves in a collisionless magnetic-field-free dusty plasma. The electrons and the ions are assumed to drift together. All the plasma components are statistically modeled by the $\kappa$-distributions with the different spectral indexes for the electrons, the ions and the dust grains. From the kinetic theory, we obtain the generalized dispersion relation, Eq.(18), and the generalized instability growth rate, Eq.(20), of the dust acoustic waves. It is shown that the dispersion relation and the instability growth rate depend on those physical quantities, such as the mass, the temperature, the density ratios of the charged particle of the components, and the dust charge state, not only, but also on the spectral indexes, $\kappa_\alpha$ ($\alpha = e, i, d$), of the different plasma components.

By using the numerical technique, we investigate the effects of the indexes, $\kappa_\alpha$ ($\alpha=e, i, d$), respectively, on the instability growth rate in the dust plasma. We find that the index $\kappa_e$ of the electrons in the plasma has a stabilized effect on the dust acoustic waves if the normalized wave number $k\lambda_{De}$ is small, but it nearly has no any effect if the normalized wave number $k\lambda_{De}$ is large. The index $\kappa_i$ of the ions in the plasma has an instability effect on the dust acoustic waves if the normalized wave number $k\lambda_{De}$ is small, but if the normalized wave number $k\lambda_{De}$ is large, it also nearly has no any effect on the rate. If the normalized wave number $k\lambda_{De}$ is small, in reverse, the index $\kappa_d$ of the dust grains in the plasma has nearly no any effect on the instability growth rate of the dust acoustic waves, but if the normalized wave number $k\lambda_{De}$ is large, the index $\kappa_d$ of the dust grains in the plasma has a stabilized effect on the dust acoustic waves. If we let $\kappa_e = \kappa_i = \kappa$ and $\kappa_d = \infty$ in Eq.(20), namely, the electrons and the ions have the same $\kappa$ index, but the dust grains be in Maxwellian distribution, the deviation of the ions and the electrons from Maxwellian distribution in this case will lead to the increase of the instability growth rate of the dust acoustic waves.

Finally, when one applies the generalized dispersion relation, Eq.(18), and the generalized instability growth rate, Eq.(20), to the dust plasma, by using the formulation for the nonextensive parameter in the nonequilibrium plasma, the nonextensive parameters $q_\alpha$ ($\alpha = i, e, d$) can be written in terms of the relation,[17]

$$k_B \nabla T_\alpha + (1-q_\alpha)Q_\alpha \nabla \phi = 0, \qquad (29)$$

where $Q_\alpha = -e$ for $\alpha = e$, $Q_\alpha = Ze$ for $\alpha = i$, and $Q_\alpha = Z_d e$ for $\alpha = d$; the potential function, $\phi = \sum_{\alpha=i,e,d} \phi_\alpha$, is determined by the Poisson's equation, Eq.(6). The three kappa indexes, $\kappa_\alpha$ ($\alpha = e, i, d$), can be calculated directly by replacing the nonextensive parameters $q_\alpha$ ($\alpha = i, e, d$) in Eq.(29) with the transformation, $\kappa_\alpha = 1/(q_\alpha - 1)$. If $q_\alpha(\alpha = i, e, d) = 1$, we have $\nabla T_\alpha = 0$, which stands for the thermal equilibrium state, where the particles are in Maxwellian distribution. But if $q_\alpha(\alpha = i, e, d) \neq 1$, we have $\nabla T_\alpha \neq 0$, which stands for the nonequilibrium stationary-state, where the particles are in non-Maxwellian distribution. They all therefore have the clear physical meanings given in the dust plasma. The reader might also be interested in some recent applications of nonextensive statistical mechanics, where a physical meaning of the parameter $q$ is introduced to astrophysical systems.[25]

## ACKNOWLEDGEMENT

This work is supported by the National Natural Science Foundation of China under the grant No.10675088.